\documentclass[prc,aps,nofootinbib,twocolumn,superscriptaddress]{revtex4}   
\usepackage{epsfig}  
\usepackage{graphicx}  
\usepackage{amssymb}
\usepackage{amsmath}
\usepackage{color}  

\usepackage{dsfont}

\begin{document}  
  
\title{ Time-dependent mean field determination of the excitation energy in transfer reactions: application to the reaction $^{238}$U on $^{12}$C at 6.14 MeV/A}

\author{G.~Scamps}  
 \email{scamps@nucl.ph.tsukuba.ac.jp}  
\affiliation{Center for Computational Sciences, University of Tsukuba, Tsukuba 305-8571, Japan }
\affiliation{Department of Physics, Tohoku University, Sendai 980-8578, Japan}

\author{C.~Rodr\'iguez-Tajes}
\affiliation{Grand Acc\'el\'erateur National d'Ions Lourds (GANIL), CEA/DRF-CNRS/IN2P3, Bd. Henri Becquerel, 14076 Caen, France}
\affiliation{Universidade de Santiago de Compostela, E-15706 Santiago de Compostela, Spain}

\author{D.~Lacroix}
\affiliation{Institut de Physique Nucl\'eaire, IN2P3-CNRS, Universit\'e Paris-Sud,
Universit\'e Paris-Saclay, F-91406 Orsay Cedex, France}

\author{F.~Farget}
\affiliation{Grand Acc\'el\'erateur National d'Ions Lourds (GANIL), CEA/DRF-CNRS/IN2P3, Bd. Henri Becquerel, 14076 Caen, France}

\date{\today}

\begin{abstract} 
The internal excitation of nuclei after multi-nucleon transfer is estimated by using the time-dependent mean-field theory.  
Transfer probabilities for each channel as well as  the energy loss after re-separation are calculated. 
By combining these two informations, we show that
the excitation energy distribution of the transfer fragments  can be obtained separately for the different transfer
channels.  The method is applied 
to the reaction involving a $^{238}$U beam on a $^{12}$C target, which has recently been measured at GANIL. It is shown that the excitation 
energy calculated  with the microscopic theory compares well with the experimental observation, provided that the competition with 
fusion is properly taken into account. The reliability of the excitation energy is further confirmed by the comparison with the phenomenological HIPSE model at higher center of mass energies.
\end{abstract}

\keywords{ }
\maketitle

\section{Introduction}
\label{intro} 
The development of a new generation of time-dependent 
mean-field codes \cite{Kim97,Nak05,Eba10,Sim12,Mar14,Bul16}, which are able to perform symmetry unrestricted 
large amplitude nuclear motion, with eventually superfluid effects, is an important step towards a unified 
microscopic description of the many facets of nuclear dynamics. 
These approaches have recently been applied to describe  a variety of phenomena,  such as collective motion \cite{Toh01,Ave08,Eba10,Ste11,Has12,Sca13,Par13,Sca14}, fusion  and transfer reactions \cite{Uma08,Sim12b,Sek13,Sca13b,Dai14, Was15,Sca15, Rei16, Has16,Vop16}, or fission \cite{Sim14,Sca15b,God15,God16,Bul16,Tan16}. Dynamical mean-field theory is particularly suitable when
 quantum aspects of single-particles are important, and the internal excitation of the system is not too 
 high. For this reason, it appears adequate for nuclear reactions close to the Coulomb barrier where fusion and transfer reactions compete.    
The time-dependent Hartree-Fock (TDHF) and TDHF+BCS theories have been used for instance to study 
dynamical effects on the fusion barriers \cite{Vop16} or dissipative transport coefficients \cite{Was09}.
By using projection techniques borrowed from nuclear structure studies, it was also possible to extract the transfer probabilities for 
1 neutron (1 proton), 2 neutrons (2 protons), ... channels and to analyze possible effects of superfluidity \cite{Sca13b}.  
Due to nucleon exchange, the transfer process also induces a global dissipation of the energy leading to an energy loss between the entrance and exit channels. 
The time-dependent mean-field theory gives access, for a given impact parameter and beam energy, to the energy loss 
averaged over the different transfer channels. This energy loss is a priori properly described because TDHF correctly 
includes the one-body dissipation effects associated to deformation and nucleon exchange processes \cite{Gro78,Blo78,Ran80,Ran84}. 
 Detailed measurements of $^{238}$U+$^{12}$C transfer reactions have been performed at GANIL \cite{Rod14,Rod16}. Data on the excitation of the transfer products have been individually obtained for each transfer channel, and the dependence with the center-of-mass energy has been experimentally investigated. These findings, which bring an experimental probe of the transfer mechanism, have largely motivated the present work.  The possibility to get any observable channel--by--channel from a 
mean-field theory is a particularly interesting and challenging problem. Indeed, TDHF or its extensions, can be seen as an average over the different channels.  Therefore, the possibility to trace-back individual probabilities instead of the average is to be clarified. 
One recent example where this was possible  concerns the determination of transfer probabilities. The use of the projection operator technique \cite{Sim11} gives access to  the individual probability of each channel. In the present article, which uses this technique as starting point, we propose a method to get the excitation energy distribution  of each channel.

\section{ TDHF+BCS calculation}
\label{TDBCS_calc}

\begin{figure}[!ht]
    \centering\includegraphics[width=\linewidth]{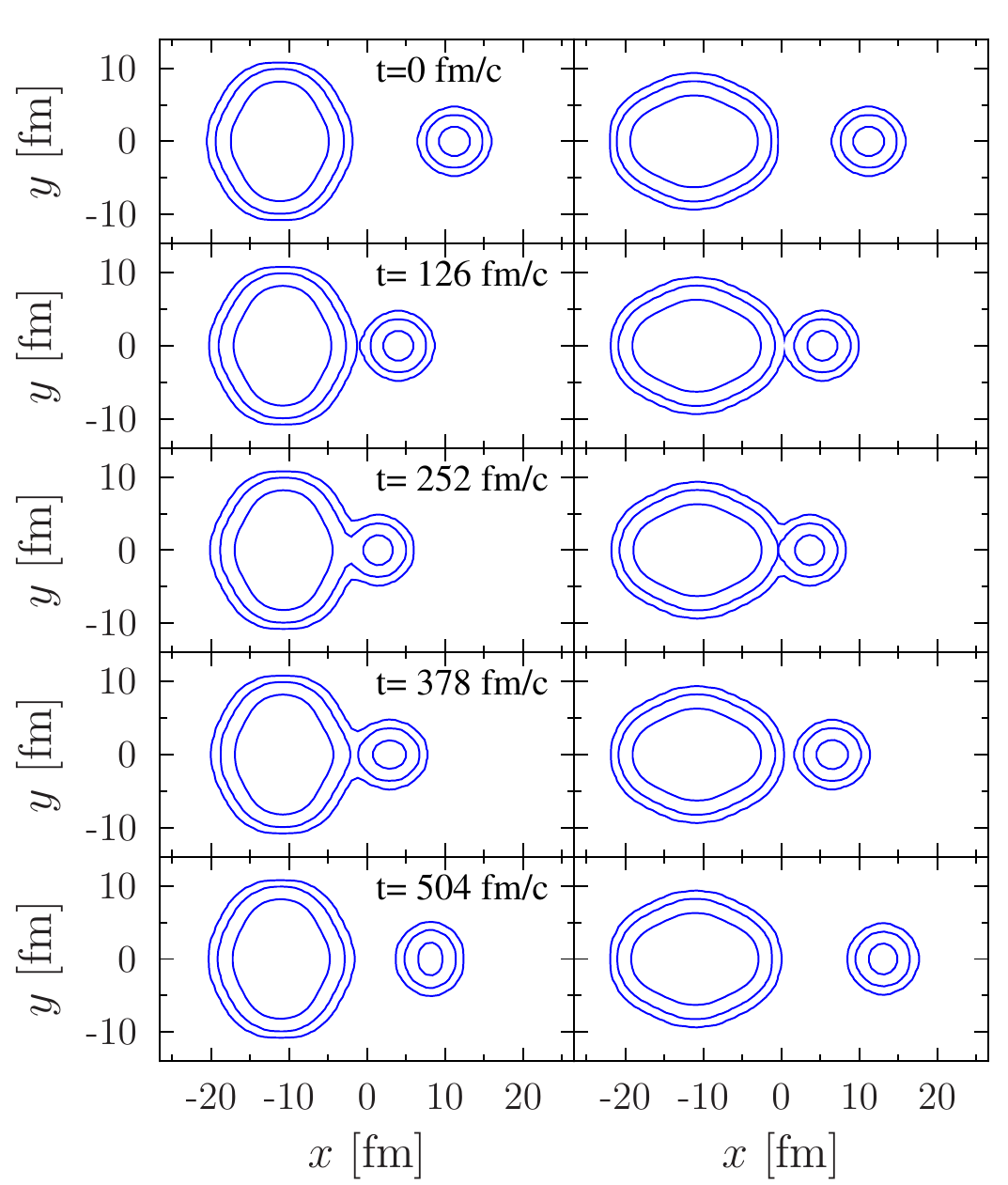}
     \caption{Snapshots of the density profile $\rho(x,y,0)$ in the center of mass at different times for the collision $^{238}$U
     on $^{12}$C. The results showed on the left- and right-side plots correspond to the perpendicular ($\theta=\pi/2$) and parallel ($\theta=0$) orientations respectively. The  center of mass energies are 
     $E'_{\rm cm} =63.75$ MeV (for $\theta=\pi/2$) and $E'_{\rm cm} = 56.25$ (for $\theta=0$).}
    \label{fig:film}
\end{figure}

We describe here the reaction $^{238}$U+$^{12}$C at 6.14 MeV/A. The $^{238}$U is superfluid and deformed in its ground state. To describe this reaction, we use the recently developed TDHF3D+BCS \cite{Sca13b} model. The pairing correlation is taken into account during the initialization of the two nuclei with the BCS approach. 
The Skyrme force Sly4d  \cite{Kim97} is used, with a surface pairing interaction \cite{Sca13b}. During the collision, the occupation numbers are assumed to be constant (frozen occupation approximation \cite{Sca12}). Calculations are done in a spatial grid of $L_x\times L_y \times 2L_z =60.8\times22.4\times22.4$ fm$^3$ with a lattice spacing $\Delta x=0.8$ fm and a time step 0.5 fm/c. Following recent 
applications of TDHF to transfer \cite{Sca13b} and 
in order to reduce the computational time,  all the calculations are done at zero impact parameter, denoted by $b$ hereafter. 
The impact parameter influence is then taken into account in an effective way by using an impact parameter dependent center of mass energy,
\begin{align}
E_{\rm cm}' &= \frac{2}{ 1 +  \sqrt{1 + \left( \frac{2b E_{\rm cm}}{Z_1 Z_2 e^2} \right)^2} } E_{\rm cm}. \label{eq:Ecm}
\end{align}
This formula is obtained by supposing that the fictitious zero impact parameter trajectory has the same minimal distance of approach as 
the  Rutherford trajectory for the impact parameter $b$ and center of mass energy $E_{\rm cm}$ considered. This approximation 
is equivalent to assume that the transfer probabilities mainly depend on the distance of closest approach \cite{Sca13b,Cor11}. 
The evolution will also depend on the angle $\theta$ between the collision axis and the orientation of the $^{238}$U. 
For each energy, two TDHF+BCS calculations are performed, at $\theta=0$ and $\theta=\pi/2$ respectively.
Examples of mean-field evolutions for energies slightly below the Coulomb  barrier are shown on Fig. \ref{fig:film}. 
We can see that the uranium is deformed in its ground state. We found a deformation parameter $\beta_2 = 0.244$.  Note that, due to deformation, the fusion barrier also depends on the orientation. By using the same method as in Ref. \cite{Sim08}, the dynamical barriers are estimated to be 57.6 MeV and 63.9 MeV, for $\theta=0$ and $\theta=\pi/2$, respectively.

\begin{figure}[!ht]
    \centering\includegraphics[width=\linewidth]{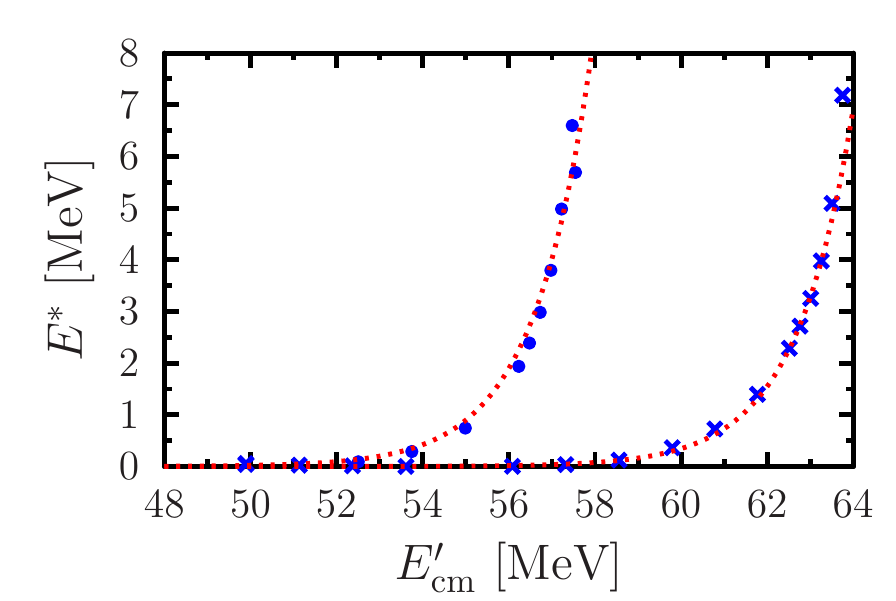}
     \caption{ Excitation energy as a function of the center of mass energy for the reaction $^{238}$U+$^{12}$C. Blue crosses and dots correspond to a perpendicular and parallel orientations of the uranium, respectively.  In both cases, the dotted red line corresponds 
     to the function \eqref{eq:exp_exc}, which adjusted to reproduce the TDHF+BCS results. }
    \label{fig:exc}
\end{figure}

For each energy $E_{\rm cm}'$, i.e. each impact parameter, the energy loss between the entrance and the exit channel is computed.
Note that during the reaction time, few light particles might be emitted. Here, we neglect the energy that might be released by 
emission and simply assume that the energy loss is entirely converted into internal excitation of the fragments. 
The excitation energy determined with this method is shown in Fig. \ref{fig:exc} as a function of the center of mass energy 
$E'_{\rm cm}$. The change of orientation shifts the function by about 6 MeV. This result is expected since the excitation comes mainly from the contact between the two fragments. The energy at which this contact occurs is lower for the parallel orientation than for the perpendicular one.

The dependence of the excitation energy with respect to the center of mass energy is well reproduced by the parametrization,
\begin{align}
E^* &= \exp \left[ a (E_{\rm cm}' - B_{\rm fus}(\theta))  + c \right], \label{eq:exp_exc}
\end{align}
with $a=0.75$ MeV$^{-1}$, and $c=1.84$. $B_{\rm fus}(\theta)$. The latter is the dynamical fusion barrier.

\begin{figure}[!ht]
    \centering\includegraphics[width=\linewidth]{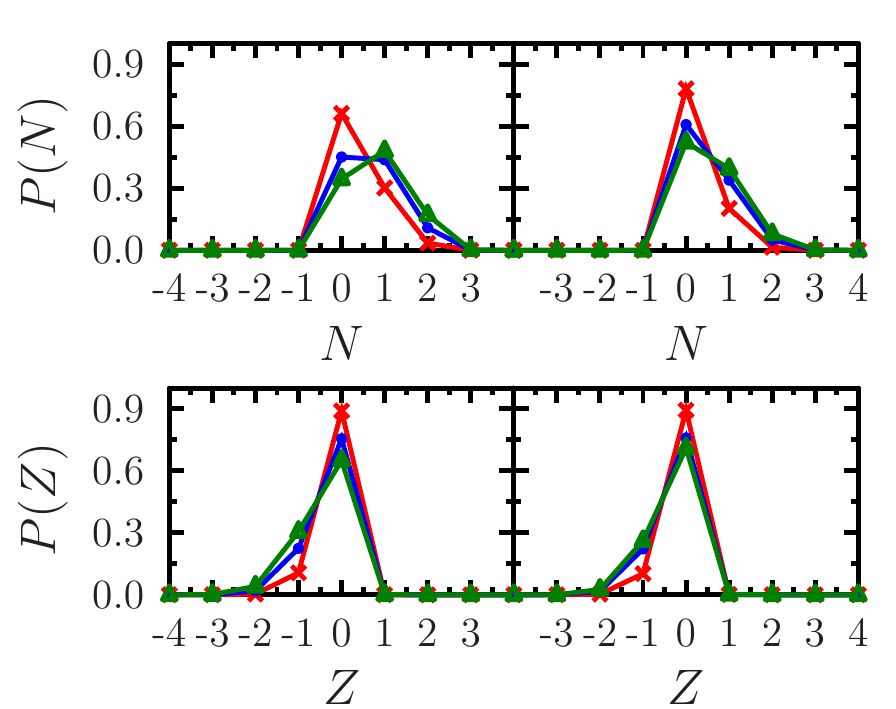}
     \caption{ Final neutron (top) and proton (bottom) transfer probability distributions obtained for the $^{238}$U+$^{12}$C reaction, for $\theta$=$\pi/2$ (left) and $\theta$=$0$ (right). The number of transferred protons and neutrons are 
     represented on the horizontal axis. Positive values correspond to transfer of nucleons from $^{238}$U to $^{12}$C.
     Calculations are done for 3 different center of mass energies: $E_{\rm cm} -B_{\rm fus}(\theta)=$ -0.1 MeV/A (green triangles), -0.4 MeV/A (blue dots) and -1.4 MeV/A (red crosses).}
    \label{fig:tr1}
\end{figure}

A second important quantity that can be extracted from the TDHF+BCS evolution is the transfer probability for different channels. 
After the collision, the probability to have $N$ neutrons and $Z$ protons in one of the sides of the lattice is computed with the projection method described in Ref. \cite{Sim11,Sca13b}. Since there is no isospin mixing, the proton and neutron numbers can be computed separately 
and we simply have $P_{\rm tr}(N,Z) = P(N) P(Z)$. Here $P(N)$ (resp. $P(Z)$) corresponds to the probability to have $N$ neutrons (resp. Z protons) after transfer. Illustrations of these probabilities are given in Fig. \ref{fig:tr1}. In particular, this figure shows the evolution of the probabilities when the center of mass energy decreases.

\begin{figure}[!ht]
    \centering\includegraphics[width=\linewidth]{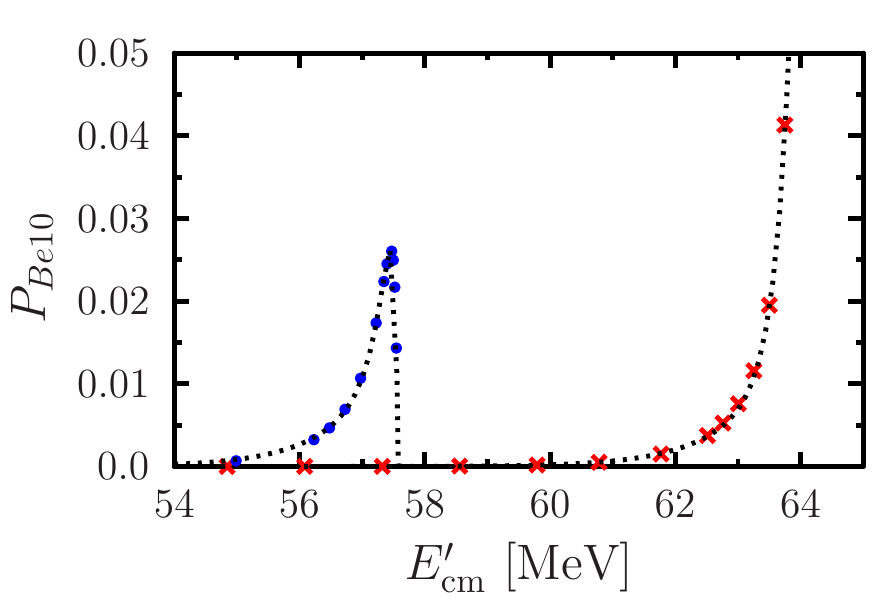}
     \caption{ Transfer probability leading to $^{10}$Be as a function of the center of mass energy for the $\theta=0$
     (blue dots) and $\theta= \pi/2$ (red crosses) orientations of $^{238}$U. 
     The black dotted lines correspond to the function \eqref{eq:trsf}, which were adjusted to reproduce the TDHF+BCS results. }
    \label{fig:trsf_2p}
\end{figure}

A systematic illustration of the probability distributions for the two considered orientations is given 
as a function of the center of mass energy for the $^{10}$Be channel in Fig. \ref{fig:trsf_2p}. 
For reasons that will be discussed below, it is convenient to parametrize the dependence observed in Fig.  \ref{fig:trsf_2p} by some analytical function. The pure transfer probabilities can be well reproduced by the formula, 
\begin{align}
P_{\rm tr}(E_{\rm cm}' )  &=  \left( e^{ a_1/E_{\rm cm}'  + c_1 } + e^{ a_2/E_{\rm cm}'  + c_2 }  \right) \nonumber \\ 
 &\times   \left( 1- e^{ a_3/E_{\rm cm}'  + c_3 } \right). \label{eq:trsf}
\end{align}
The values of the $\{a_1,c_1,a_2,c_2, a_3, c_3\}$ parameters, which were obtained by fitting the calculated transfer probabilities, 
are reported in Table \ref{tab:exp_exc}, for the different channels discussed below and the two considered orientations.

\begin{table}[!h]
\begin{tabular}{|c|c|c|c|c|c|c|}
 \hline
   $\theta= \pi/2$ & $a_1$ & $c_1$ & $a_2$ & $c_2$ & $a_3$ & $c_3$  \\
 \hline
    $^{13}$C+$^{237}$U  & -1059 &	15.57 &  -3659 & 55.69 & -3399 & 51.05 \\
  \hline  
    $^{14}$C+$^{236}$U  &  -2925 &	43.43 & -12468 & 193.3 & -12468 & 193.3 \\
  \hline  
    $^{11}$B+$^{239}$Np  &  -1822 & 26.74 & -8527 & 132.2 & -8527 & 132.2 \\
  \hline  
    $^{9}$Be+$^{241}$Pu  & -3994 &	48.83 & -24545 & 374.4 & -24545 & 374.4  \\
  \hline  
    $^{10}$Be+$^{240}$Pu  & -4222 & 61.9	 & -18126 & 280.7 & -18126 & 280.7  \\
    \hline
    \hline
$\theta=0$& $a_1$ & $c_1$ & $a_2$ & $c_2$ & $a_3$ & $c_3$  \\
 \hline
    $^{13}$C+$^{237}$U  & -1848 & 31.17 & -8783 & 149.4 & -8783 & 149.4  \\
  \hline  
    $^{14}$C+$^{236}$U  &  -3029 & 49.37 & -9133 & 155.9 & -9133 & 155.9  \\
  \hline  
    $^{11}$B+$^{239}$Np  & -2474 & 41.7 & -31598 & 548 & -26496 & 460.1 \\
  \hline  
    $^{9}$Be+$^{241}$Pu  & -3935 & 	58.73 & -62838 & 1083 & -62838 & 1083  \\
  \hline  
    $^{10}$Be+$^{240}$Pu  & -5289 & 88.27 & -28725 & 496.4 & -19225 & 333.9 \\
  \hline  
\end{tabular}
\caption{Values of the parameters entering in Eq.  \eqref{eq:trsf}, which were obtained by fitting the TDHF+BCS transfer probabilities for the main channels and the perpendicular and parallel orientations. The values of the $a_i$ parameters are given in MeV.   }
\label{tab:exp_exc}
\end{table}
From the parameterization of the excitation energy given above 
and the transfer probabilities and following the method of Ref. \cite{Sek13, Son15}, 
we can now determine the transfer cross section as a function of the excitation energy $E^*$,
\begin{align}
&\sigma_{\rm tr}(N,Z,E^*,\theta,E_{\rm cm}) = 2 \pi b(E^*, \theta, E_{\rm cm}  ) \nonumber \\
&\hspace*{2.cm}\times P_{\rm tr}(N,Z,E^*, \theta,E_{\rm cm }) \left| \frac{\partial b}{ \partial E^*} \right|.
\label{eq.sigma}
\end{align}
Eq.~\eqref{eq:Ecm} together with Eq.\eqref{eq:exp_exc} gives a bijection between the impact parameter and the excitation energy. It is therefore possible to obtain the impact parameter $b(E^*, E_{\rm cm} , \theta )$ as a function of the excitation energy $E^*$ for a given center of mass energy and a given orientation of the heavy nucleus. 

\section{Comparison with  the experiments}

\label{Experiment}

One drawback of the time-dependent mean-field approach is that it does not describe properly the possible competition
between transfer channels and fusion. Indeed, for energies above the fusion barrier, the mean-field evolution always leads to fusion while, for energies below, 
only transfer channels are populated. This limitation stems from the semiclassical nature of  the mean-field dynamics in collective space. 
Experimentally, both reaction mechanisms coexist in the vicinity of the Coulomb 
barrier. In the experiment performed at GANIL, only transfer reactions leading to fission 
have been detected. The details of the experiment can be found in Ref \cite{Rod16}. In order to compare with experimental observations, one should start from the calculated 
pure transfer probabilities $\sigma_{\rm tr}$ and account for: (i) the competition with fusion channel, and (ii) the fact that, even if transfer 
occurs, it does not automatically lead to the fission of the heavy fragment. One should then estimate in some approximate way  
the fusion and fission probabilities. 
 \begin{figure}[!ht] 
	\centering\includegraphics[width=\linewidth]{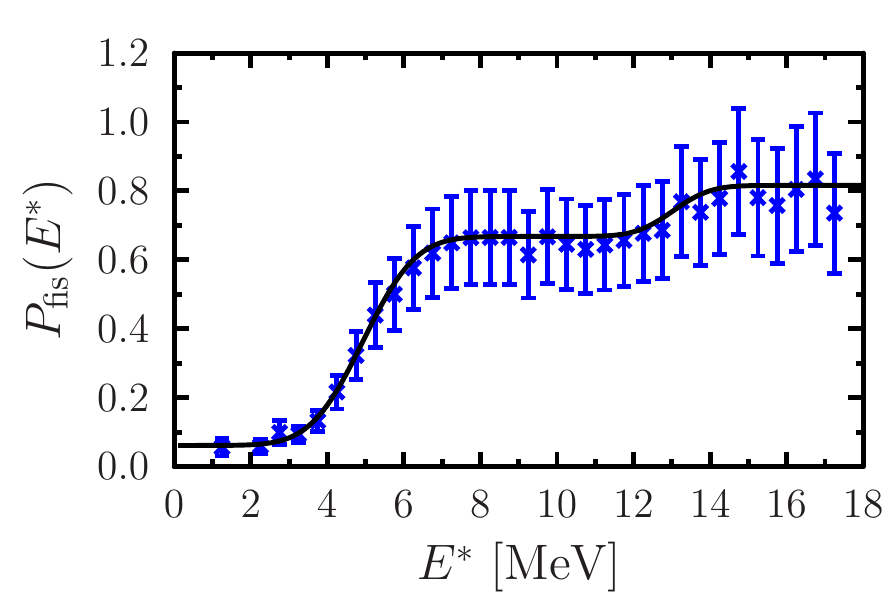}  
	\caption{ Experimental fission probability of the  $^{240}$Pu as a function of the excitation energy (blue cross) \cite{Rod14}. The solid line is fitted on the experimental data with the Eq. \eqref{eq:fit_fct_fiss}. } 
	\label{fig:Pf_240Pu} 
\end{figure} 
Here, the fusion probability is simply assumed to follow an error function \cite{Swi05},
\begin{align}
P_{\rm fus}(E_{\rm cm}',\theta) = \frac{ 1}{2} \left\{ 1+{\rm erf} \left( \frac{E_{\rm cm}'-B_{\rm fus}(\theta)}{\sigma} \right) \right\},
\end{align}
where dynamical fusion barriers $B_{\rm fus}(\theta)$ deduced from TDHF+BCS for the two considered orientations are used.
The fluctuation of the barrier height is assumed to be $\sigma$=0.5 MeV. 
Note that the fusion probabilities for all possible orientations are necessary. A method will be discussed below that smoothly 
interpolates between a parallel and a perpendicular orientation of the heavy nucleus.

\begin{figure}[!ht]
	\centering\includegraphics[width=\linewidth]{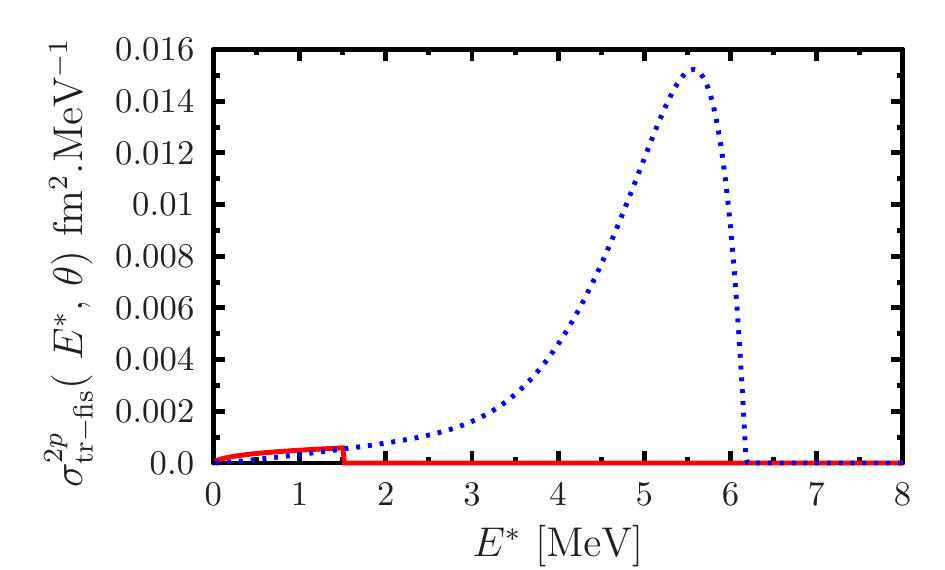}  
	\caption{  Transfer-induced fission cross section for the reaction $^{12}$C($^{238}$U,$^{240}$Pu)$^{10}$Be as a function of the excitation energy  for  the two orientations of $^{238}$U, $\theta$=0 (blue dotted line) and $\theta$=$\pi/2$ (red solid line).  Here, the simulated center of mass energy $E_{cm} = 6.14$ MeV/A is the same for the two orientations.} 
	\label{fig:D_orient} 
\end{figure}

For the fission probability, we took advantage of the experimental results obtained in Ref. \cite{Rod14} (see Fig. \ref{fig:Pf_240Pu}), where the fission probability has been obtained as a function of excitation energy. The experimental probability is well reproduced by the parameterization,
\begin{align}
	  P_{\rm fis}(E^*) = f_1 + f_2 \; {\rm erf} ( f_3 E^* - f_4  ) + f_5 \; {\rm erf} ( f_6 E^* - f_7), \label{eq:fit_fct_fiss}
\end{align}
with $f_1$=0.4384, $f_2$=0.3034, $f_3$=0.6554~MeV$^{-1}$, $f_4$=3.233, $f_5$=0.07403, $f_6$=0.9167~MeV$^{-1}$ and $f_7$=11.94 for the  $^{240}$Pu.  
Results of the this parametrization is shown in Fig. \ref{fig:Pf_240Pu} with solid line. Note that different transfer channels 
lead to different heavy systems that might eventually fission. For simplicity, we assume here that all these heavy systems have a fission probability 
equal to the one measured for $^{240}$Pu.  The impact of this approximation has been investigated through specific calculations, which took into account the different fission probability of each system. The results, which only showed minor effects, justified such a simplification.
Having now phenomenological prescriptions for the fusion and fission probabilities, we can deduce the transfer-induced fission probability.
This probability is calculated as the pure transfer cross-section convoluted by the probabilities that no fusion occurred and that the heavy system did fission. It is therefore computed as,
\begin{align}
\sigma_{\rm tr,fis}(N,Z,E^*,\theta,E_{\rm cm} ) &= \sigma_{\rm tr}(N,Z,E^*,\theta,E_{\rm cm} ) \nonumber \\
&\times \left( 1-P_{\rm fus}(E_{\rm cm}',\theta) \right) P_{\rm fis}(E^*). \label{eq:crosstffis}
\end{align}
An illustration of this cross section is displayed in Fig. \ref{fig:D_orient} for the two-proton transfer channel from $^{12}$C, leading to $^{10}$Be. We see that the orientation affects considerably the cross-section. Indeed, for a given center of mass energy,  
for the parallel orientation, the two fragments come in contact leading to rather high excitation while,  for the perpendicular case, 
the two fragments remain well separated at all time and are much less excited.

\begin{figure}[!ht]
	\centering\includegraphics[width=\linewidth]{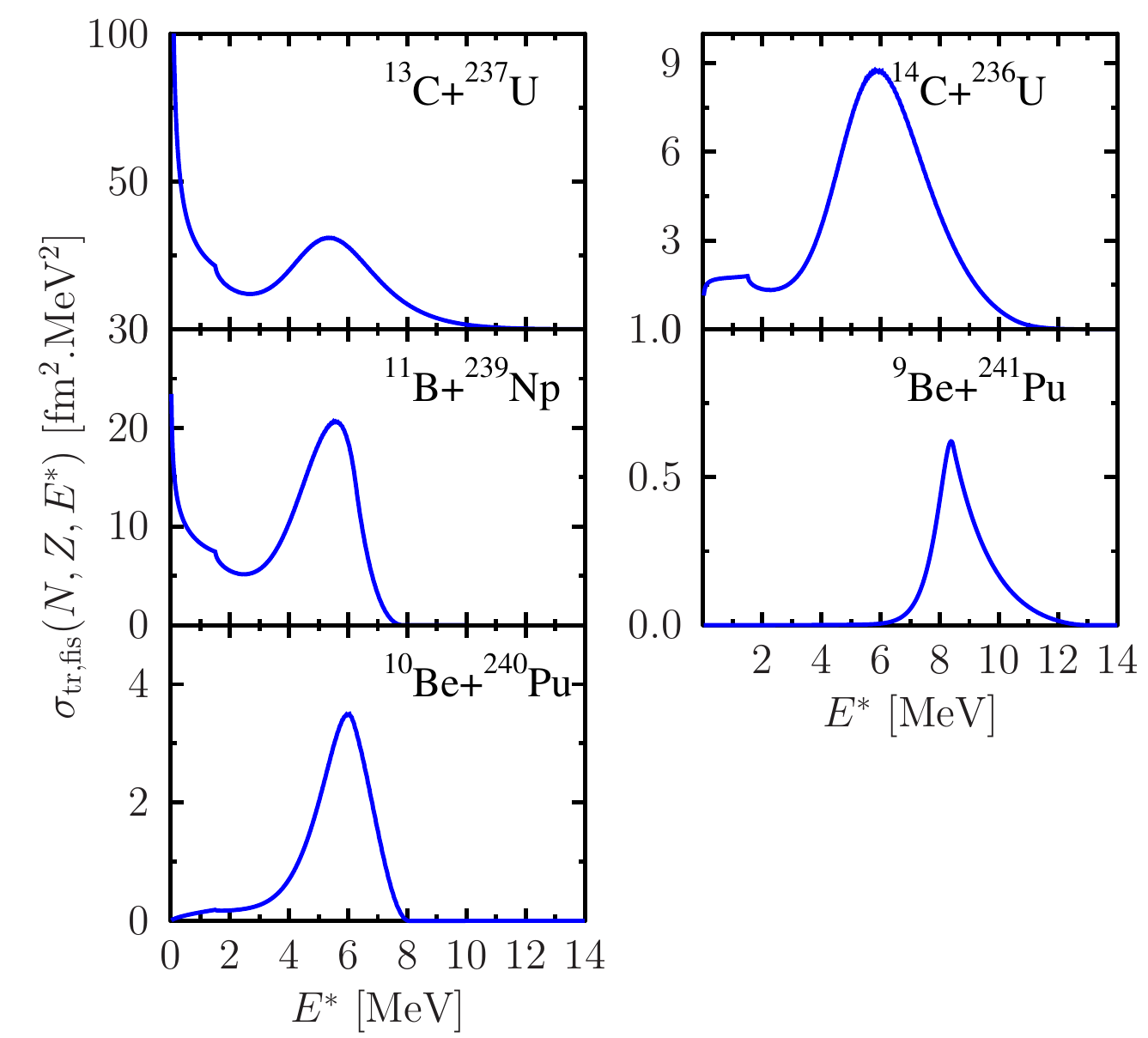}  
	\caption{$\sigma_{\rm tr,fis}(N,Z,E^*)$ as a function of excitation energy for the main transfer 
	channels observed experimentally at a center of mass energy of 59 MeV.} 
	\label{fig:crosstrfiss} 
\end{figure}

In order to obtain a cross section comparable to experiment, one should 
average equation (\ref{eq:crosstffis}) on all possible orientations:
\begin{align}
\sigma_{\rm tr,fis}(N,Z,E^*,E_{\rm cm} ) &= \nonumber  \\
 4 \int_0^{\pi/2} \sigma_{\rm tr,fis}&(N,Z,E^*,\theta,E_{\rm cm} )  \sin(\theta) d \theta . \label{eq:final}
\end{align}
Here, the cross-section for all possible angles between $0$ and $\pi/2$ is obtained by simple interpolation of different parameters 
entering in the Eqs. (\ref{eq:exp_exc}) and (\ref{eq:trsf}).
 This interpolation is done in a linear way for ${\cal O}=B_{\rm fus}$ or $\sigma_{\rm tr}$,
\begin{align}
{\cal O}(\theta) &= z {\cal O}(\theta=\pi/2) + (1.-z)  {\cal O}(\theta=0),
\end{align}
and exponentially for $a_1$, $a_2$, $a_3$, $c_1$, $c_2$ and $c_3$,
\begin{align}
{\cal O}(\theta) &= {\cal O}(\theta=\pi/2)^z {\cal O}(\theta=0)^{(1.-z)},
\end{align}
with $z$=$\sin(\theta)^2$. Using this interpolation method, the  cross section is integrated as a function of all possible 
orientations of the $^{238}$U. The obtained cross sections are shown in fig. \ref{fig:crosstrfiss} for the main channels.

 \begin{figure}[!ht]
	\centering\includegraphics[width=\linewidth]{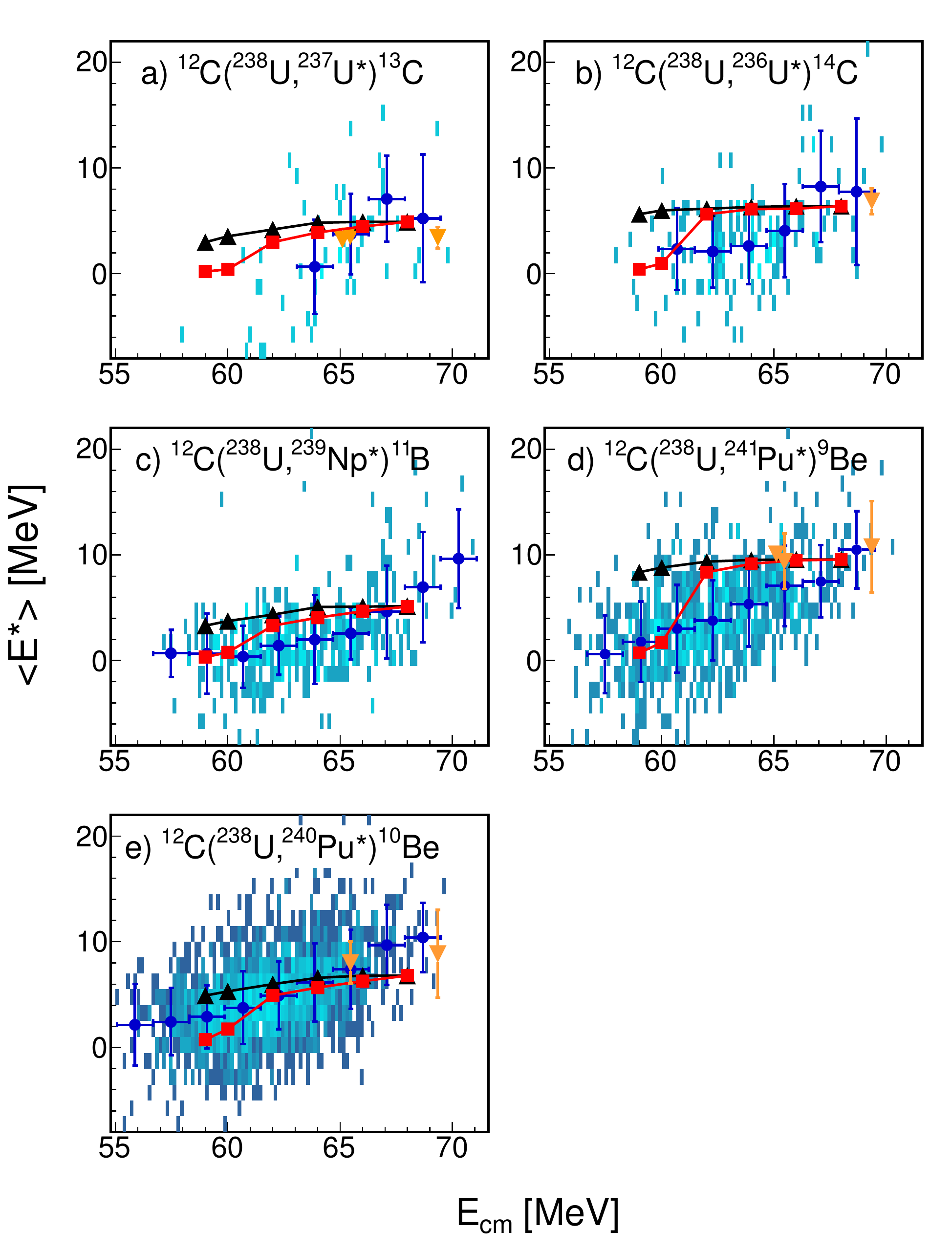}  
	\caption{ Average excitation energy as a function of the center of mass energy for the main channels 
	observed experimentally: experimental data (blue dots), TDHF+BCS results (black triangles), TDHF+BCS results where the excitation energy have been shifted by 3 MeV (red squares) and the HIPSE results (orange down triangles). In the latter case, error bars corresponds to the widths of the calculated distribution. The superimposed blue areas correspond to the experimental event--by--event distributions of the excitation energy. } 
	\label{fig:aver_all_comp_shift_bar} 
\end{figure} 
From the cross section $\sigma_{\rm tr,fis}(N,Z,E^*)$,  we can compute the average value of the excitation energy as a function of the center of mass energy for each transfer channels. These values are compared with the experimental results in Fig. \ref{fig:aver_all_comp_shift_bar}. 
While the TDHF+BCS results are in relatively good agreement with the data at high center of mass energies, a systematic overestimation 
is seen in the low energy regions. The main reason for this discrepancy is the difference between the barrier height obtained by TDHF+BCS (62 MeV) and the experimental one (65 MeV). 
This difference is most probably due to the poor treatment of light systems like $^{12}$C in the mean-field approach. 
In order to correct for this systematic error, 
we have artificially shifted the average excitation energy used in the cross-section calculation with TDHF by 3 MeV  
, i.e
\begin{align}
\langle E^* \rangle_{\rm cor}(E_{\rm cm}) = \langle E^* \rangle(E_{\rm cm} - 3 {\rm MeV}).
\end{align}
With this correction, the general behavior of the average excitation energy as a function of the center of mass energy is in better agreement with 
the experimental results. In particular, for the $^{10}$Be channel.

\section{HIPSE calculations}


In this section, a different theoretical approach will be discussed, the Heavy-Ion Phase-Space (HIPSE) model
developed in Ref. \cite{Lac04} which was originally designed for heavy-ion reactions around the Fermi energy, and it is able to describe
a variety of reaction mechanisms including secondary decay leading eventually to fission. The main ingredient of this approach is the 
exploration of all the accessible phase-space configurations at the minimal distance of approach, accounting for all the energetic and geometric constraints. See Ref. \cite{Lac04} for further details. Due to the semi-classical treatment of the incoming nuclei, it could only be used above 
the fusion barrier and therefore it can be seen as a complementary study compared to TDHF+BCS. A set of events has been generated with HIPSE 
for several beam energies: $5.6$ MeV/A, $5.7$ MeV/A,   $5.8$ MeV/A,  $6.0$ MeV/A and $6.14$ MeV/A. In each case, $8.10^{6}$ events have been 
generated with impact parameters ranging from $0$ to $4$ fm. Note that above 4 fm, the two nuclei do not touch each other due to the high charge 
of the Uranium beam. The different simulations have been performed using an adiabatic parameter equal to -0.13 and a transfer rate equal to $60\%$ while it is assumed that no direct two-body collisions occurs due to the Pauli blocking effect. Note that changing slightly the two former parameters does not affect the result. One advantage of the HIPSE model is that it gives access to the fragment partition as well as their internal 
excitation before decay. Using this information, we have estimated the average excitation energy of the heavy fragment before secondary decay 
for the different channels populated in the experiments of Refs. \cite{Rod14,Rod16}. Results are shown by orange triangles in Fig. \ref{fig:aver_all_comp_shift_bar}. When no results are shown, it means that the channel was not populated by HIPSE. Note that, this model 
being optimized at higher beam energies, it is not expected to be predictive for the transfer cross-sections. However, 
we see that, when 
the channel is observed in HIPSE, the average excitation energy is rather close  from both  the experimental and TDHF+BCS estimates. This gives further confidence in the values obtained here using the microscopic mean-field approach. 
 
\section{Summary} 

In the present work, a method is proposed to determine the evolution of the excitation energy as a function of the center of mass energy with the TDHF+BCS theory. All the calculations being done by assuming a head-on collision, a method is used in order to effectively take into account the impact parameter effects. The deformation is taken into account via an interpolation method between the perpendicular and parallel orientations of the heavy nucleus.
To make contact with experiments, several ingredients should be taken into account.
In particular, the fusion as well as the fission  probabilities should be used to transform the pure transfer probabilities 
obtained with our microscopic theory into a transfer--fission probability that was measured experimentally. 
It is shown that the microscopic theory 
slightly overestimates the experimental excitation energy at low center of mass energy (most peripheral collisions) while at higher excitation energy, a good agreement is obtained. This discrepancy can be traced back in the poor treatment of very light systems like $^{12}$C in the mean-field 
approach. Once this aspect is corrected,  the experimental results for the reaction $^{238}$U+$^{12}$C compares 
rather well with TDHF+BCS calculations. In particular, the evolution of the excitation energy as a function of the center of mass for the main channel. The mean-field and experimental values of excitation energy at high center of mass energy are also consistent with the HIPSE results. 

While the time-dependent mean-field approach does a priori give average information on the nuclear reaction, 
the present work further confirms that the mean-field theory can be also helpful to extract chemical or energetic information 
on transfer reaction channel--by--channel.
For future applications, calculations taking into account explicitly the impact parameter \cite{Sek13} could be done in order to improve the results. Another improvement can be achieved using the TDHFB theory instead of TDHF+BCS with Skyrme \cite{Ste11} or Gogny effective interaction \cite{Has16}.

\begin{acknowledgments}
G.~S. acknowledges the Japan Society for the Promotion of Science 
for the JSPS postdoctoral fellowship for foreign researchers. 
This work was supported by Grant-in-Aid for JSPS Fellows No. 14F04769. 
This project has received funding from
the European Unions Horizon 2020 research and innovation
program under grant agreement No. 654002. This work has also been supported by a postodoctoral grant from the Regional Council of Galicia, within the \textit{Plan galego de investigaci\'on, innovaci\'on e crecemento 2011-2015}.
\end{acknowledgments}

\end{document}